\documentstyle[12pt]{article}

\newcommand{\wt}{\widetilde}

\def\la{\mathrel{\mathpalette\fun <}}
\def\ga{\mathrel{\mathpalette\fun >}}
\def\fun#1#2{\lower3.6pt\vbox{\baselineskip0pt\lineskip.9pt
\ialign{$\mathsurround=0pt#1\hfil##\hfil$\crcr#2\crcr\sim\crcr}}}

\begin{document}

\title{VORTICES ON THE HIGGS BRANCH OF THE SEIBERG--WITTEN
THEORY }

\author{Alexei Yung\\
{\em Petersburg Nuclear Physics Institute}\\
{\em Gatchina, St.Petersburg 188350, Russia}\\
E-mail: yung@thd.pnpi.spb.ru}
\date{June 1999}
\maketitle

\begin{abstract}

We study the mechanism of confinement via formation of
Abrikosov--Nielsen--Olesen vortices on the Higgs branch of $N=2$
supersymmetric SU(2) gauge theory with  massive fundamental
matter. Higgs branch represents a limiting case of
superconductor of type I with vanishing Higgs mass. We show that
in this limit vortices still exist although they become
logarithmically "thick". Because  of this the confining
potential is not linear any longer. It behaves as $L/\log L$
with a distance $L$ between confining heavy charges (monopoles).
This new confining regime can occur only in supersymmetric
theories. We also address the problem of quantum stability of
vortices. To this end we develop string representation for a
vortex and use it to argue that vortices remain stable.
\end{abstract}
\vspace*{1cm}
PNPI-TH-2319

\newpage

\section{Introduction}

Supersymmetric gauge theories can be viewed as a "theoretical
laboratory" to develop insights in the dynamics of strongly
coupled gauge theories like QCD. Revolutionary ideas of
electromagnetic duality \cite{MO,SW1} made it possible to study
traditionally untractable strongly coupled theories using weak
coupling description in terms of dual variables. Particularly
spectacular results were obtained by Seiberg and Witten in $N=2$
supersymmetry where the low energy effective Lagrangians were
found exactly \cite{SW1,SW2}.

One of the most important physical outcome of the
Seiberg--Witten theory is the demonstration of U(1) confinement
via monopole condensation \cite{SW1}. According to old
Mandelstam and t'Hooft ideas \cite{M} when monopoles condense
the electric flux is confined in the (dual)
Abrikosov--Nielsen--Olesen (ANO) vortex tubes \cite{ANO}
connecting heavy trial charge--anticharge pair. Because this
vortex has a constant energy per length (string tension $\tau$)
the potential between charge and anticharge increases linearly
with their separation. Thus, if monopoles (charges) condense
then charges (monopoles) confine. The effective description of
this phenomenon is given by the Abelian Higgs model.

The ratio of the Higgs mass $m_H$ to the photon mass $m_\gamma$
in the Abelian Higgs model is an important parameter
characterizing the type of the superconductor. For
$m_H>m_\gamma$ we have type II superconductor while for
$m_H<m_H$  we have type I.

In the Seiberg-Witten theory this confinement scenario is
realized in two possible ways. First, in pure gauge theory near
monopole (dyon) singularity upon breaking $N=2$ supersymmetry to
$N=1$ one by adding the mass term to the adjoint matter
\cite{SW1}. Second, in $N=2$ theory with fundamental matter
hypermultiplets on Higgs branches \cite{SW2}.

Consider the $N=2$, SU(2) gauge theory \cite{SW1}. The gauge
group is broken down to U(1) by the expectation value
$\langle\varphi^a\rangle=\delta^{a3}a$ of the adjoint scalar
field $\varphi^a$, $a=1,2,3$. The complex parameter $a$
parametrize the modular space (Coulomb branch) of the theory.
Near the monopole point on the Coulomb branch (the point, where
monopole becomes massless) the effective low energy theory is
the dual $N=2$ QED. This means that the theory has light
monopoles coupled to dual photon in the same way as ordinary
charges are coupled to the  ordinary photon.

Upon breaking $N=2$ supersymmetry by the mass term for the
adjoint matter monopoles condense implying the U(1) confinement
for heavy trial charges. The effective low energy theory is
given by the Abelian Higgs model with $m_H=m_\gamma$ \cite{SPA}
(here $m_H$ is the mass of the monopole and $m_\gamma$ is the
mass of the dual photon). This condition means that the ANO
vortex saturate the Bogomolny bound \cite{B}
(see \cite{ENS} for the relation between supersymmetry  and the
BPS bound) and the string tension $\tau$ is given by
\begin{equation} \tau\ =\ 2\pi v^2 \end{equation}
at least for small masses of the adjoint matter.
\footnote{It is argued in \cite{HSZ} that this string is not in
fact the BPS-\-saturated state (it does not belong to the
"short"  multiplet) due to higher corrections in the mass
of the adjoint matter.} Here $v$ is the VEV of the Higgs field.

The second option is the confinement on Higgs branches in the
Seiberg--Witten theory. If two (or more) flavours of matter of
the $N=2$ SU(2) gauge theory with $N_f$ fundamental
hypermultiplets have the same mass then the Higgs branch is
developed with $\langle Q\rangle\neq0$, where $Q$ is the matter
hypermultiplet \cite{SW2}. Positions of roots of Higgs
branches on the Coulomb branch depend on the mass parameter $m$.
The low energy theory near a given singularity (root) is $N=2$
Abelian Higgs model, in which the role of the Higgs
hypermultiplet is played by the state becoming massless in this
singularity.

Once charges (monopoles) condense on the Higgs branch we expect
monopoles (charges) to confine \cite{SW2,HSZ}. However, this
scenario is much less understood, than the first one discussed
above. The purpose of the present paper is to study the
mechanism of the confinement on Higgs branches via formation of
ANO vortices. We consider the case of two hypermultiplets $N_f=2$
with the common mass $m$ for simplicity.

From the point of view of the low energy Abelian Higgs  model
the Higgs branch is a limiting case of the superconductor of
type I with $m_H=0$.  The usual arena for studies of ANO
vortices is the opposite, so called London limit of type II
superconductivity with $m_H\gg m_\gamma$. In the London limit
the string tension is given by
\begin{equation}
\tau_{II}\ =\ 2\pi v^2\ln\frac{m_H}{m_\gamma}\ .
\end{equation}
This result was obtained by Abrikosov in 1957 and then rederived
by Nielsen and Olesen in the framework of the relativistic field
theory \cite{ANO}.

In this paper we study vortices in the limit $m_H=0$. Classical
vortex solutions still exist for the type I superconductor
\cite{BV} and we show that they survive the limit $m_H=0$.
However, their transverse size becomes logarithmically large. As
a result the string tension becomes suppressed as
\begin{equation}
\tau\ =\  \frac{2\pi v^2}{\ln m_\gamma L}\ ,
\end{equation}
where $L$ is the length of the string (the distance between
heavy monopole and antimonopole). In fact, $\tau$ in (1.3) goes
to zero for infinite strings in accordance with \cite{R}.

Our result (1.3) means quite unusual behavior from the point of
view of confinement. Eq.(1.3) means that the confining potential
between well separated heavy sources (monopole and antimonopole
in the case at hand) is not linear in $L$ any longer. It behaves
as
\begin{equation}
V(L)\ =\ 2\pi v^2\ \frac L{\ln m_\gamma L}
\end{equation}
 for large $L$.

This peculiar behavior is specific for supersymmetric gauge
theories with Higgs branches (we don't have Higgs branches with
$m_H=0$ without supersymmetry). The potential (1.4) is the order
parameter which distinguish this new confinement phase from the
ordinary confinement phase with $V(L)\sim L$ or from other
phases. The potential (1.4) gives rise to the following
behavior of the Wilson loop produced by the infinitely heavy
monopole going along the contour $C$. For the
$W^D_C=P\exp i\int_Cdx_\mu A^D_\mu$ ($A^D_\mu$ is the dual
potential) the obvious generalization of (1.4) gives
\begin{equation}
W^D_C\ =\ \exp\left\{-\frac{2\pi^2v^2S}{\ln(m_\gamma\cdot2S/P)}
\right\}
\end{equation}
for large $S$ and $P$, where $S$ is the area spanned by the loop
$C$ and $P$ is its perimeter. For rectangular loops $S=TL$ and
$P=2(L+T)$, where $T$ is the time interval. We see that (1.5) is
different from the usual area law for the Wilson loop.

The paper is organized as follows. In Sect.2 we review the Higgs
branches in the $N_f=2$ Seiberg--Witten theory and derive the low
energy action on the Higgs branch. In Sect.3 we study classical
vortex solution and derive the result (1.3). Then in Sect.4 we
address the problem of quantum stability of vortices. We develop
string representation for the vortex and use it to argue that
vortices survive in the quantum theory.
 Sect. 5 contains our conclusions.

\section{Review of Higgs branches \newline in SU(2) theory}
\setcounter{equation}{0}

Let us introduce $N_f=2$ fundamental matter hypermultiplets in
the $N=2$ SU(2) gauge theory. In terms of $N=1$ superfields
matter dependent part of the microscopic action looks like
\begin{eqnarray}
&& S_{\rm matter}=\int d^4xd^2\theta d^2\bar\theta\left[\bar Q_A
e^VQ^A+\bar{\wt  Q}^Ae^V\wt Q_A\right]\ + \nonumber\\
+&& i\int d^4xd^2\theta\left[\sqrt2\wt Q_A\frac{\tau^a}2\
Q^A\Phi^a+m\wt Q_AQ^A\right]+\mbox{ c.c.}
\end{eqnarray}
Here $Q^{kA},\wt Q_{Ak}$ are matter chiral fields, $k=1,2$ and
$A=1,\ldots,N_F$, while $V$ is the vector superfield. $\Phi^a$
is the adjoint chiral superfield (its scalar component
$\varphi^a$ develops VEV $a$). Thus we have 16 real matter
degrees of freedom for $N_f=2$.

Consider first the limit of large $m$, $m\gg\Lambda$. Then three
singularities on the Coulomb branch are easy to understand. Two
of them correspond to monopole and dyon singularities of the
pure gauge theory. Their positions on the Coulomb branch are
given by \cite{SW2}
\begin{equation}
u_{m,d}\ =\ \pm\ 2m\Lambda-\frac12\Lambda^2\ ,
\end{equation}
where $u=\frac12\langle\varphi^{a^2}\rangle\simeq a^2/2.\ $
\footnote{We use the Pauli-Villars regularization scheme for the
normalization of $\Lambda$, see \cite{FP}.} In the large $m$
limit $u_{m,d}$ are approximately given by their values in the
pure gauge theory $u_{m,d}\simeq\pm2m\Lambda=\pm2\Lambda^2_0$,
where $\Lambda_0$ is the scale of $N_f=0$ theory.

The third singularity corresponds to the point where charge
becomes massless. Let us decompose matter fields as
\begin{equation} Q^{kA}\
=\ \left({1\atop 0}\right)^k Q^A_++\left({0\atop 1}
\right)^kQ^A_-\ .
\end{equation}
From the superpotential in (2.1) we see that the $Q_+$ becomes
massless at
\begin{equation}
a\ =\ -\ \sqrt2\ m\ .
\end{equation}
The singular point $a=+\sqrt2\,m$ is gauge equivalent to the one
in (2.4). In terms of variable $u$ (2.4) reads
\begin{equation}
u_c\ =\ m^2+\frac12\ \Lambda^2\ .
\end{equation}
Strictly speaking, we have $2+N_f=4$ singularities on the
Coulomb branch. However two of them  coincides for the case of
two flavors of matter with the same mass.

The effective theory on the Coulomb branch near charge
singularity (2.4) is given by $N=2$ QED with light matter fields
$Q^A_+$, $\wt Q_{+A}$ (8 real degrees of freedom) as well as the
photon multiplet.

The charge singularity (2.4),(2.5) is the root of the Higgs
branch \cite{SW2}. To find this branch let us write down $D$-term
and $F$-term conditions which follow from (2.1). $D$-term
conditions are
\begin{equation}
Q^{kA}\bar Q_{A\ell}+\bar{\wt Q}^{kA}\wt Q_{A\ell}\ =\ 0\ ,
\end{equation}
while $F$-term conditions give (2.4) as well as
\begin{equation}
Q^{kA}\wt Q_{A\ell}\ =\ 0\ .
\end{equation}
Eqs. (2.6),(2.7) have nontrivial solutions for $N_f\ge2$. These
solutions determines VEV's for scalar components  $q^{kA}$, $\wt
q_{Ak}$ of fields $Q^{kA}, \wt Q_{Ak}$. Dropping heavy
components $q_-$ according to decomposition (2.3) and
introducing the SU$_R(2)$ doublet $q^{fA}$ as
\begin{eqnarray}
q^{1A}=\ q^A_+\ , && q^{2A}=\ \bar{\wt q}^A_+\ , \nonumber\\
\bar q_{A1}=\ \bar q^+_A\ , && \bar q_{A2}=\ -\wt q^+_A\ ,
\end{eqnarray}
we can rewrite three real conditions in (2.6),(2.7) as
\begin{equation}
\bar q_{Ap}(\tau^a)^p_f\ q^{fA}\ =\ 0, \quad a=1,2,3.
\end{equation}
Eq.(2.9) together with the condition (2.4) determines the Higgs
branch (manifold with $\langle q\rangle\neq0$) which touches the
Coulomb branch at the point (2.4).

The low energy theory for boson fields near the root of the Higgs
branch looks like
\begin{equation}
S^{\rm root}_{\rm boson}=\int d^4x\left\{\frac1{4g^2}
F^2_{\mu\nu}+\bar\nabla_\mu\bar q_{Af}\nabla_\mu q^{fA}+
\frac{g^2}8[\mbox{ Tr }\bar q\tau^aq)]^2\right\},
\end{equation}
where trace is calculated over flavor and SU$_R(2)$ indices.
Here $\nabla_\mu=\partial_\mu-in_eA_\mu$, $\bar\nabla_\mu=
\partial_\mu+in_eA_\mu$, the electric charge $n_e=1/2$ for
fundamental matter fields.

This is an Abelian Higgs model with last interaction term coming
from the elimination of $D$ and $F$ terms. The QED coupling
constant $g^2$ is small near the root of Higgs branch (we
specify its value later on). We include 8 real matter degrees of
freedom $q^{fA}$ in the theory (2.10) according to the
identification (2.8). The rest of matter fields $q^A_-$, $\wt
q\,^-_A$ (another 8 real degrees of freedom) acquire large mass
$2m$ and can be dropped out. The effective theory (2.10) is
correct on the Coulomb branch near the root of the Higgs branch
(2.4) or on the Higgs branch not far away from the origin
$\langle q\rangle=0$.

It is clear that the last term in (2.10) is zero on the fields
$q$ which satisfy constraint (2.9). This means that moduli
fields which develop VEV's on the Higgs branch are massless, as
it should be. The other fields acquire mass of order
$\langle\bar qq\rangle^{1/2}$. It turns out that there are four
real moduli fields $q$ (out of 8) which satisfy the constraint
(2.9) \cite{SW2}. They correspond to lowest components of the one
hypermultiplet.

We can parametrize them as
\begin{equation}
q^{f\dot A}(x)\ =\ \frac1{\sqrt2}\ \sigma^{f\dot A}_\alpha
\phi_\alpha(x) e^{i\alpha(x)}\ .
\end{equation}
Here $\phi_\alpha(x), \alpha=1\ldots4$ are four real moduli
fields. It is clear that fields (2.11) solve (2.9). The common
phase $\alpha(x)$ in (2.11) is the U(1) gauge phase. Once
$\langle\phi_\alpha\rangle=v_\alpha\neq0$ on the Higgs branch the U(1)
group is broken and $\alpha(x)$ is eaten by the Higgs  mechanism.
Say, in the unitary gauge $\alpha(x)=0$. In the next section we
consider vortex solution for the model (2.10). Then $\alpha(x)$
is determined by the behavior of the gauge field at the
infinity. Substituting (2.11) into (2.10) we get the bosonic
part of the effective theory on the Higgs branch near the origin
\begin{equation}
S^{\rm Higgs}_{\rm boson}\ =\ \int d^4x\left\{\frac1{4g^2}\
F^2_{\mu\nu}+\bar\nabla_\mu\bar q_\alpha\nabla_\mu
q_\alpha\right\}\ ,
\end{equation}
where
\begin{equation}
q_\alpha(x)\ =\ \phi_\alpha(x)\ e^{i\alpha(x)}\ .
\end{equation}

Once $v_\alpha\neq0$ we expect monopoles (they are heavy at
$m\gg\Lambda$) to confine via formation of vortices which carry
the magnetic flux. In the rest of the paper we study in detail
vortex solutions in the model (2.12). As we mentioned in the
Introduction this model corresponds to the type  I
superconductivity with the vanishing Higgs mass, $m_H=0$. The
photon mass in the model (2.12) is
\begin{equation} m^2_\gamma\ =\ \frac12\ g^2v^2_\alpha\ .
\end{equation}

Note, that we do not introduce
Fayet--Illiopoulos term in the Seiberg--Witten theory. In the
presence of such term there are BPS-\-saturated vortices. They
correspond to special points on the modular space where VEV's of
massless moduli fields $\langle q_\alpha\rangle=0$, whereas
other massive matter fields from (2.10) develop VEV's equal to
Fayet--Illiopoulos parameters with windings at spatial infinity
\cite{AHISS,HSZ}.  Instead, vortices which arise in theories
with flat Higgs potential are not BPS-\-saturated. They are
discussed from the brane point of view in \cite{V} and from the
field theory point of view in \cite{R}.

If we increase $v^2_\alpha$ taking $v^2_\alpha\ga\Lambda^2$ we
can integrate out massive photon. Then the effective theory is a
$\sigma$-model for massless fields $q_\alpha$ which belong to
4-dimensional Hyper--Kahler manifold, $R^4/Z_2$. The metric of
this $\sigma$-model is flat \cite{SW2,APS}, there are, however,
higher derivative corrections induced by instantons \cite{Y}. In
this paper we consider region of Higgs branch with
$v^2_\alpha\ll\Lambda$. This determines the scale of the
effective Abelian Higgs model (2.12). $W$-bosons and other
particles which reflect the non-\-Abelian structure of the
underlying microscopic theory are heavy with masses $\ga\Lambda$
and can be ignored.

To conclude this section let us briefly review what happens if
we reduce the mass parameter $m$. At $m=\pm\Lambda$ the charge
singularity (root of the Higgs branch) collides with the monopole
(dyon) singularity, see Eqs.(2.2),(2.5). These are
Argyres--Douglas points \cite{AD,APSW}. At these points mutually
non-\-local degrees of freedom (say, charges and monopoles)
becomes massless simultaneously. These points are very
interesting from the point of view of the monopole confinement on
the Higgs branch we study in this paper. Mono\-pol\-es become
dynamical as we approach Argyres--Douglas point, $m\to\Lambda$.
We comment on the physics at this point at the end of the paper.

After the collision quantum numbers of particles at
singularities change because of monodromies \cite{SW2}. If we
denote quantum numbers as $(n_m,n_e)_B$, where $n_m$ and $n_e$
are magnetic and electric charges of the state, while $B$ is its
baryon number then at $m>\Lambda$ we have charge, monopole and
dyon singularities with quantum numbers
\begin{equation}
(0,\ 1/2)^{*2}_1\ , \quad (1\ 0)_0\ , \quad (1,\ 1)_0\ .
\end{equation}
The superscript for the charge means that we have two flavors
of charges. After charge singularity collides with monopole one
(at $m<\Lambda)$ the quantum numbers of particles at
singularities become \cite{BF}
\begin{equation}
(1,\ 0)^{*2}_0\ , \quad (1,\ 1/2)_1\ , \quad (1,\ 1/2)_{-1}\ .
\end{equation}
Now monopole $(1,0)_0$ condense on the Higgs branch which
emerges from the point (2.5), while dyons $(1,1/2)_1$ and
$(1,1/2)_{-1}$ confine because they carry electric
charge.\footnote{Throughout the rest of the paper to avoid
confusion we refer to particles which develop VEV's on the Higgs
branch as charges and to states which confined as monopoles
(dyons). This terminology corresponds to masses of matter above
Argyres--Douglas point $m>\Lambda$. At $m<\Lambda$ one should
change the terminology according to Eqs. (2.15),(2.16).} At
zero mass, $m=0$ two dyon singularities in (2.16) coincide (see
(2.2)) and the second Higgs branch appears at the point
$u=-1/2\,\Lambda^2$. This restores the global symmetry from
SU$(N_f=2)$ in the massive theory to SO$(2N_f=4)$ at $m=0\ $
\cite{SW2}.

\section{Vortex solution}
\setcounter{equation}{0}

In this section we construct the classical vortex solution for
the model (2.12). Without loss of generality we can rotate VEV's
$v_\alpha$ in the $R^4$ space to the configuration
$v_\alpha=(v,0,0,0)$. Now we look for vortex solutions with
$q_1\equiv q\neq0$ and $q_2=q_3=q_4=0$. Thus for the purpose of
finding the classical vortex solution we can drop fields
$q_2,q_3,q_4$ in (2.12) \footnote{We cannot ignore these
massless fields in the quantum theory, however.}. Doing so we
arrive at the standard Abelian Higgs model with one complex
matter field $q\equiv q_1$
\begin{equation}
S_{AH}\ =\ \int
d^4x\left\{\frac1{4g^2}\ F^2_{\mu\nu}+ \bar\nabla_\mu\bar
q\nabla_\mu q+\lambda(|q|^2-v^2)^2\right\}.
\end{equation}
Here we write
down the double well potential term for the matter field $q$
with coupling constant $\lambda$. We cannot introduce this term
in the Seiberg--Witten theory without breaking down $N=2$
supersymmetry. We use this term to regularize the vortex
solution and afterwards take the limit $\lambda\to0$. In terms
of $\lambda$ the mass of the Higgs field reads
\begin{equation}
m^2_H\ =\ 4\lambda\,v^2\ .
\end{equation}
We consider the model (3.1) at
\begin{equation}
m_H\ \ll\ m_\gamma\ ,
\end{equation}
(see (2.14)) taking the limit $m_H\to0$ in the end. The limit
(3.3) is the opposite to the London one.

\subsection{Two dimensional problem}

To begin with in this subsection we study infinitely long ANO
vortex, with length $L\gg1/m_H$. In fact, this is not what we
want. To take limit $m_H\to0$ we have to study not that long
vortices. We focus on the opposite limit $L\ll1/m_H$ in the next
 subsection.

For infinitely long vortex the problem becomes two dimensional.
Assume that solutions for $q$ and $A_\mu$ depend only on $x_i$,
$i=1,2$. Let us use the standard ansatz for fields $q$ \cite{B}
\begin{equation}
q(x)\ =\ \phi(r)\ e^{i\arg(x^i)}\ ,
\end{equation}
where $\phi(r)$ is real, and for $A_\mu$ \begin{equation} n_eA_i(x)\ =\
-\varepsilon_{ij}\ \frac{x_j}{x^2}[1-f(r)]\ , \end{equation} where
$r^2=x^2_i$, while $A_0=A_3=0$. We also assume standard boundary
conditions for $\phi(r)$
\begin{eqnarray}
&& \phi(\infty)\ =\ v\ , \nonumber\\
&& \phi(0)\ =\ 0\ ,
\end{eqnarray}
and
\begin{eqnarray}
&& f(0)\ =\ 1\ , \nonumber\\
&& f(\infty)\ =\ 0
\end{eqnarray}
for $f(r)$.

The boundary conditions (3.7) for gauge field ensures that the
vortex carry one unit of the magnetic flux. To see this let us
calculate the dual field strength
\begin{equation}
F^*\ =\ \varepsilon_{ij}\partial_iA_j\ =\ -\frac1{n_er}\ f'(r)\
.
\end{equation}
Then using (3.7) one gets
\begin{equation}
n_e\int d^2xF^*\ =\ 2\pi\ .
\end{equation}
Substituting (3.4), (3.5) into the action (3.1) we get the
energy of the vortex as a function of $\phi$ and $f$
\begin{equation}
\varepsilon\ =\ 2\pi L\int dr\left\{\frac1{2n_eg^2} \frac{f'^2}r
+r\phi'^2+\frac{f^2}r\phi^2+\lambda(\phi^2-v^2)^{2}r\right\}.
\end{equation}
The equations of motion read for field $\phi$
\begin{equation}
\phi''+\frac1r\phi'-\frac1{r^2}f^2\phi-m^2_H
\frac{\phi(\phi^2-v^2)}{2v^2}\ =\ 0
\end{equation}
and for $f$
\begin{equation}
f''-\frac1rf'-\frac{m^2_\gamma}{v^2}\phi^2f\ =\ 0\ .
\end{equation}

Now to solve these equations in the limit $m_\gamma\gg m_H$ we
adopt the following model for the vortex. We assume that the
electromagnetic field of the vortex is confined inside the core
of radius $R_g$ (we find $R_g$ later), such that $F^*=0$ at
$r\ge R_g$. Instead  scalar field $\phi$ is almost zero inside
the core and grows towards its  VEV $\phi=v$ at the infinity
outside the core.

Thus at zero order approximation  we assume that $\phi=0$ at
$r\le R_g$. We improve this approximation later on and show that
corrections to $\phi$ inside the core do not affect the string
energy in the logarithmic approximation.

If $\phi=0$ inside the core we have from Eq.(3.12) for gauge
field $f$
\begin{equation}
f^{(0)}\ =\ \left[\begin{array}{cl}
\displaystyle 1-\frac{r^2}{R^2_g}\ , & r\le R_g\\
\displaystyle 0\ , & r\ge R_g\ . \end{array}
\right.
\end{equation}
The superscript (0) means the zero order approximation. To this
order the magnetic field is constant inside the core and zero
outside, see (3.8).

Now consider Eq.(3.11) for $\phi(r)$ outside the core.  Taking
into account (3.13) and neglecting the nonlinearity in (3.11) at
large $r\ $ \footnote{As soon as we are going to take the limit
$m_H\to0$ the last term in (3.11) plays really the role of a
mass term which regularize the behavior of $\phi$ at
$r\to\infty$ and the nonlinearity associated with this term is
not important.} we find that (3.11) reduces to the equation of
motion of free field with mass $m_H$. We  have at $r\gg R_g$
\begin{equation}
(\phi-v)\ \sim\ -vK_0(m_Hr)\ .
\end{equation}
Here $K_0$ is a Bessel function. It has exponential fall-off at
infinity  $K_0(m_Hr)\sim e^{-m_Hr}$ and logarithmic behavior at
small $m_Hr$:
\begin{equation}
K_0(m_Hr)\ \simeq\ \ln\frac2{m_Hr}\ .
\end{equation}
We assume that $R_g\ll1/m_H$ (we confirm this later) so
$K_0(m_Hr)$ becomes very large at $r\sim R_g$. To match the
solution (3.14) with the value $\phi=0$ inside the core we have
to normalize (3.14) properly. Doing so we finally get for the
field $\phi$
\begin{equation}
\phi^{(0)}\ =\ \left[ \begin{array}{cl}
\displaystyle 0\ , & r\le R_g\\
\displaystyle v\left(1-\frac{K_0(m_Hr)}{\ln(2/m_HR_g)}\right), &
r\gg R_g. \end{array} \right.
\end{equation}
Substituting (3.13) and (3.16) into (3.10) we get the energy of
vortex as a function of $R_g$
\begin{equation}
\varepsilon\ =\ 2\pi L\left\{\frac8{g^2}\frac1{R^2_g}
+\frac{v^2}{\ln(2/m_HR_g)}\left[1+0\left( \frac1{\ln(1/m_HR_g)}
\right)\right]\right\}.
\end{equation}
The first term here comes from the electromagnetic energy inside
the core (see the first term in (3.10)). The second term comes
from the scalar field "surface energy" (the second term
in(3.10)). All other terms are small in powers of
$1/\ln(m_HR_g)$. Note, that the second term in (3.17) comes from
the logarithmic integration in the region $R_g\ll r\ll1/m_H$.
This logarithm in the numerator cancels one of two logarithms in
the denominator (coming from the normalization in (3.16)) giving
the result (3.17).

Minimizing (3.17) with respect to $R_g$ we find
\begin{equation}
R^2_g\ \simeq\ \frac8{m^2_\gamma}\ \ln^2\frac{m_\gamma}{m_H}\ .
\end{equation}
Substituting this into (3.17) we get finally  the vortex energy
\begin{equation}
\varepsilon_0\ =\ \frac{2\pi v^2}{\ln(m_\gamma/m_H)}\ L\ .
\end{equation}
We stress once again that the main contribution to the vortex
energy (3.19) comes from the logarithmic "tail" (3.16) of the
scalar field outside the core.

So far we have considered
vortices with winding number $n=1$. Now it is clear that to the
leading order the vortex energy does not depend on $n$ for
$m_H\ll m_\gamma$ and is given by Eq.(3.19).

We can calculate corrections to our zero order approximation
vortex (3.13),(3.16) in the intermediate region $r\sim R_g$. For
example, taking into account (3.13) in Eq.(3.11) we get the
improved behavior for $\phi(r)$ inside the core $r\la R_g$
\begin{equation}
\phi(r)\ =\ \frac v{\ln(2/m_HR_g)+1}\ \frac r{R_g}\ +\ \cdots\ ,
\end{equation}
where dots stand for higher powers of $r/R_g$.

In a similar way we can get the next-to-leading correction to
$f(r)$ at small $r$
\begin{equation}
f(r)\ =\ 1-\frac{r^2}{R^2_g}+\frac18\ \frac{r^4}{R^4_g}\ +\
\cdots\ .
\end{equation}
It is easy to see, however, that these corrections do not affect
the main contribution to the energy of the vortex (3.19) in the
logarithmic approximation.

Now let us discuss the result (3.18),(3.19). If we formally take
the limit $m_H\to0$ then the vortex becomes infinitely "thick"
($R_g\to\infty)$ and its string tension goes to zero. This
result is in the accordance with results of Ref.\cite{R}, where
it was noted that there are no vortices for the theories with
flat Higgs potential. The reason is that field $\phi$ has
logarithmic behavior outside the core and cannot approach its
boundary condition $\phi=v$ at infinity. This can be seen from
(3.16). If we put the IR-\-regularization $m_H$ to zero in
(3.16) then $K_0(m_Hr)$ is given by (3.15) and $\phi$ goes to
infinity instead of $v$ at $r\to\infty$.

It is clear therefore, that the infinitely long vortex does not
exist at $m_H=0~~$ \footnote{See, however, the discussion of
tensionless strings in the brane picture \cite{HK}.} In the next
subsection we consider vortices of the finite length $L$ and
show that $L$ plays the role of IR-regularization in the limit
$m_H=0$.

\subsection{Three dimensional problem}

Now, we are going to consider long but not infinitely long
vortices $1/m_\gamma\ll L\ll1/m_H$. This will allow us to take
the limit $m_H\to0$.

As we already mentioned before the main problem with the limit
$m_H\to0$ is that in this limit the field $\phi$ outside the
core obeys free field equation (see (3.11)). Thus $\phi\sim\ln
r$ at large $r$ and cannot approach its boundary value $v$ at
$r\to\infty$.

However, it is clear that for the vortex of finite length $L$ at
$|x|\gg L$ field $\phi$ behaves as $\phi-v\sim1/|x|$. Thus it
can reach its boundary value at infinity.

Let us consider the vortex of length $L$ stretched in $x_3$
direction between static heavy monopole at the point $x_3=L/2,$
$x_1=x_2=0$ and antimonopole at $x_3=-L/2$, $x_1=x_2=0$. More
specifically we consider the same ansatz (3.4),(3.5) for fields
$q$ and $A_\mu$ with winding phase $\alpha(x)=\arg(x_1,x_2)$ for
$|x_3|\ll L$ and $\alpha=0$ for $|x_3|\gg L$. Functions $\phi$
and $f$ now depend on $r^2=x^2_1+x^2_2$ and $x_3$.

The electromagnetic field is confined in the narrow tube of
radius $R_g$ (we determine $R_g$ later) which is stretched
between monopole and antimonopole. We assume the same behavior
for $f(r)$ as in (3.13) at $|x_3|\ll L$ and put $f=0$ at $|x_3|\gg
L$.  For field $\phi$ we have $\phi=0$ at $r<R_g$ (with possible
corrections like in (3.20)), $\phi-v\sim\ln r/L$ at $R_g\ll r\ll
L$ and $(\phi-v)\sim-1/|x|$ at $|x|\gg L$.

Normalizing these solutions to fit them together in the
intermediate regions $r\sim R_g$ and $|x|\sim L$ we get
\begin{equation}
\phi^{(0)}=\left\{\begin{array}{l}
\displaystyle 0\ ,\quad  r\le R_g\ ,\quad  x_3\ll L\\
\displaystyle v-\frac v{\ln L/R_g}\ln\frac Lr\ ,\ \;  R_g\ll
r\ll L,\ \;  x_3\ll L\\
\displaystyle  v-c\frac v{\ln L/R_g}\frac L{|x|}e^{-m_H|x|},\
\quad |x|\gg L\ , \end{array} \right.
\end{equation}
where constant $c\sim1$.

Substituting (3.13),(3.22) into the action (3.1) we get the
vortex energy as a function of $R_g$
\begin{equation}
\varepsilon\ =\ 2\pi
L\left\{\frac8{g^2}\frac1{R^2_g}+\frac{v^2}{\ln(L/R_g)}
\left[1+0\left(\frac1{\ln(L/R_g)}\right)\right]\right\}.
\end{equation}
Here the first term comes from the electromagnetic energy in
(3.1) and the second one from the kinetic energy of field $\phi$
("surface energy") in the region of logarithmic integration
$R_g\ll r\ll L$. Minimizing (3.23) with respect to $R_g$ we get
\begin{equation}
R^2_g\ =\ \frac8{m^2_\gamma}\ln^2m_\gamma L
\end{equation}
and the vortex energy
\begin{equation}
\varepsilon_0\ =\ 2\pi v^2\ \frac L{\ln m_\gamma L}\ ,
\end{equation}
which gives our result (1.3) for the string tension $\tau=
\varepsilon_0/L$.

Again it is easy to see that corrections to functions $f$ and
$\phi$ in the intermediate regions $\gamma\sim R_g$ and $|x|\sim
L$ do not change (3.25). Moreover, we can make our model for the
vortex more complicated. Introduce another free parameter $R_0$
(besides $R_g$) and write down
\begin{equation}
\phi\ =\ v-\frac v{\ln(R_0/R_g)}\ln\frac{R_0}r\ ,
\end{equation}
at $R_g\ll r\ll L$. This region gives the main
contribution to the energy. Minimizing the energy with respect
to $R_0$ we find $R_0\sim L$ and the vortex energy is still
given by (3.25).

Our results in (3.24), (3.25) show that the vortex length $L$
really plays the role of the IR-cutoff. Now we can safely put
$m_H=0$.  The radius of the vortex remains finite and its energy
nonzero.

The result in (3.25) means that monopoles are in the confinement
phase on the Higgs branch of the Seiberg--Witten theory.
However, the confinement potential is not linear in $L$. It
shows weaker behavior $L/\log L$ at large $L$. This new
confinement phase is a specific one for supersymmetric theories
with Higgs branches.

Of course, results of this section are purely classical. In the
next section we address the  problem of quantum stability of
vortices on the Higgs branch.

\section{Quantum stability of vortices}
\setcounter{equation}{0}

\subsection{General considerations}

In a given Higgs vacuum with $\langle\bar qq\rangle\neq0$ the
stability of vortices is ensured by the topology. Vortices
correspond to minimum energy configurations with a given winding
number $n$, which is an element of $\pi_1(U(1))=Z.~~$
\footnote{As we have shown in the last section the vortex
energy does not depend on winding number $n$ in the limit
$m_H=0$.  Thus vortex with winding number $n$ has less energy
than  $n$ vortices with $n=1$. We discuss this issue in the
subsection 4.3.}

However, one might suspect that the presence of vortices can
destabilize the Higgs branch causing the phase transition to the
unbroken phase with $\langle\bar qq\rangle=0$. The reason for
this is that in the theory with flat Higgs potential there is no
loss of volume energy associated with transition from the broken
phase with $\langle\bar qq\rangle\neq0$ to the unbroken phase
with $\langle\bar qq\rangle=0$.

Let us compare two configurations. First is the
monopole--antimonopole pair at large separation $L$ in the Higgs
vacuum $\langle\bar qq\rangle=v^2\neq0$. The magnetic flux of
the monopole creates the vortex string connecting the monopole
with the antimonopole. The energy of this configuration is given
by (3.25). Second, consider the same monopole--antimonopole pair
in the unbroken vacuum with $\langle\bar qq\rangle=0$.   The
energy of this configuration associated with Coulomb interaction
of monopole and antimonopole is
\begin{equation}
\varepsilon\ \sim\ \frac1L\ ,
\end{equation}
with possible logarithmic corrections coming from the
renormalization of the coupling constant.

Comparing (3.25) and (4.1) at large $L$ one can see that the
monopole--antimonopole pair in the unbroken vacuum has less
energy. Therefore, one may suspect that the presence of vortices
induces the collapse of the Higgs branch with the phase
transition to the Coulomb branch with $\langle\bar qq\rangle=0$.

Now it is clear that the problem is whether the  barrier of this
transition is finite or infinite. We argue in this section that
the barrier is in fact infinite and the Higgs branch remains
stable.

Another possible way to understand our conclusion about the
stability of vortices we are going to make in this section is
to note that the estimate (4.1) is hardly correct.
The point is that once the "electric" coupling constant
$g^2$ is small then the "magnetic" one is large. This means
that there are uncontrollable corrections to the Coulomb
interactions of the monopole and the antimonopole in (4.1).

Of course, we cannot prove that the vortex is stable in the quantum
theory (still, we present some general arguments in subsection 4.3).
However, in this section we study several natural possibilities
for the vortex to develop instability and show that they do not
lead to the collapse of the Higgs branch.

First, imagine that the radius of the vortex $R_g$ grows and at
large $R_g$ the vortex becomes a bubble of the unbroken vacuum
inside the broken one (remember that $q\simeq0$ inside the
vortex core).

From (3.23) we see that the energy increases with $R_g$ as
$(\log L/R_g)^{-1}$ at $R_g\ll L$. Moreover, it is easy to
estimate the surface energy of vortex if $R_g\gg L$. It goes as
$v^2R_g$, thus infinite $R_g$ means infinite barrier.

Another possibility is that we have many bubbles of finite size
which coalesce producing the unbroken vacuum. This means that we
have many strings on the Higgs branch (they should be closed if
monopoles are heavy). We discuss this possibility in subsection
4.3.

In the next subsection we study one more possibility for the
string to develop instability. Imagine that the string develops
large fluctuations actually covering the whole space. This may
lead to the transition to the unbroken phase. To study this
possibility we develop the string representation for the vortex.

\subsection{The string representation}

In this subsection we are going to find out if fluctuations of
vortex string (deviation of vortex from a straight line) are
controlled by any small parameter or not. To this end we develop
string representation for the vortex. To do this we follow the
logic of Refs.\cite{O,SY}, where the string representation for
vortex in the London limit $(m_H\gg m_\gamma)$ has been
developed. See also recent paper \cite{BS} where the string
representation for the case $m_H=m_\gamma$ is suggested.

To work out the string representation for the vortex we integrate
out photon multiplet and use $m_\gamma$ $(R^{-1}_g$ in (3.24),
more exactly) as a UV-\-cutoff for our string theory. This leads
to the $\sigma$-model for the real matter field $\phi(x)$ (see
(3.4)) with the flat metric and the constraint $\phi=0$ on the
string worldsheet $z_\mu(\sigma)\ $ \footnote{We also ignore
$\phi_2$, $\phi_3$ and $\phi_4$ in (2.13) because they decouple
from $\phi=\phi_1$ to the leading order.}. We have for the
partition function for  the boson sector of the theory
\begin{eqnarray}
Z &=&\int D\phi(x)Dz_\mu(\sigma)D\rho(\sigma)J(\phi)
\quad \times \\
&\times&\exp\left\{-\int d^4x\left[(\partial_\mu\phi)^2+\int
d^2\sigma\,\delta^{(4)}
(x-z(\sigma))i\rho(\sigma)\phi(x)\right]\right\}.\nonumber
\end{eqnarray}
Here we introduce field $\rho(\sigma)$ living on the string
worldsheet as a Lagrange multiplier.
 The Jacobian $J(\phi)$ is defined by
\begin{equation}
J(\phi)\int Dz_\mu(\sigma)\prod_\sigma\delta(\phi(z))\ =\ 1\ .
\end{equation}
The Jacobian $J(\phi)$ is important for the quantization of the
string theory. As it is argued in \cite{PS,ACPZ} its role is to
make the ANO string critical in four dimensions. We do not work
out $J(\phi)$ here. It involves quantum corrections in the
coupling constant of the $2D$ $\sigma$-model, which appears to
be small as we show below.

Writing down $\phi=v+\delta\phi$ and integrating over
$\delta\phi$ in (4.2) we get the $2D$ theory on the string
worldsheet
\begin{eqnarray}
Z &=& \int Dz_\mu D\rho\exp-\left\{\int d^2\sigma
iv\rho(\sigma)\quad + \right. \nonumber\\
&+& \left. \int d^2\sigma d^2\sigma'\frac1{4(2\pi)^2}
\frac{\rho(\sigma)\rho(\sigma')}{[z(\sigma)-z(\sigma')]^2}
\right\}.
\end{eqnarray}
The theory in (4.4) is a non-local one. It contains all powers
of derivatives of $z_\mu(\sigma)$. To work out the leading term
in derivatives let us fix the gauge
\begin{equation}
z_0\ = \ \sigma_1\ , \qquad z_3\ =\ \sigma_2
\end{equation}
and consider deviations from the flat string surface $z_i$,
$i=1,2$ to be small. We have
\begin{equation}
[z(\sigma)-z(\sigma')]^2\ =\ (\sigma-\sigma')^2+
[(\sigma-\sigma')_n\partial_nz_i]^2+\cdots\ ,
\end{equation}
where $n=1,2$ runs over two coordinates in (4.5). Substituting
(4.6) into (4.4) we get the integral
\begin{equation}
\frac1{(2\pi)^2}\int\frac{d^2(\sigma-\sigma')}{(\sigma
-\sigma')^2}\ =\ \frac1{2\pi}\ln m_\gamma L\ ,
\end{equation}
where we use $m_\gamma$ as a UV-cutoff and characteristic size
(length) of the string $L$ as a IR-cutoff. Using (4.7) and
integrating over $\rho$ in (4.4) we get finally
\begin{equation}
Z\ = \int Dz_\mu\exp-\left\{\frac{2\pi v^2}{\ln m_\gamma L}
\left[TL+\int d^2\sigma\frac12(\partial_nz^i)^2+\cdots
\right]\right\}.
\end{equation}
The string tension in (4.8) coincides with the one in (1.3)
 which we
have obtained in the last section solving classical equations of
motion for the vortex (see (3.25)).

The action in the exponent in (4.8) is the expansion of
Nambu--Goto action
\begin{equation}
S_{\rm string}\ =\ \tau\int d^2\sigma\sqrt g\ ,
\end{equation}
where $g_{mn}=\partial_mz_\mu\partial_nz_\mu$ is the induced
metric and $\tau$ is given by (1.3). Corrections to (4.9)
include higher derivative terms like rigidity term \cite{P} and
so on. However the expansion in derivatives hardly makes sense
here because for $m_H=0$ the length of the string by itself
plays the role of a IR-cutoff.

Now let us see whether our string is stable or it can develop
large fluctuations. To this end let us estimate the renormalized
string tension. We have
\begin{equation}
\tau_{ren}\ =\ \tau-\frac{\rm const}{R^2_g}\ ,
\end{equation}
where we use that the size of a core $R_g$ (3.24) plays the role
of UV-cutoff for our string theory (4.9) ((4.4) more precisely).

Now let us show that the second term in (4.10) is small as
compared to the first one. Its ratio to $\tau$ plays the role of
the effective coupling constant in the $2D$ $\sigma$-model (4.9),
(4.4)

\begin{equation}
g^2_{2D}\ \sim\ \frac1{R^2_g\tau}\ \sim\ \frac{g^2}{\ln m_\gamma
L}\ .
\end{equation}
To estimate (4.11) let us recall our general strategy. We start
with the microscopic non-Abelian Seiberg--Witten theory. Then we
integrate over scales $\mu\ga\Lambda$ and end up with effective
$N=2$ QED (see (2.12)) near the root of the Higgs branch. The
QED description is valid at scales $\mu$ in the region
$m_\gamma\ll\mu\ll\Lambda$. The QED coupling constant $g^2$ is
$1/g^2\sim-\ln\mu/\Lambda_{QED}$, where
$\Lambda_{QED}=m^3/\Lambda^2$ at $m\ga\Lambda$. At the scale
$\mu=m_\gamma$ the photon multiplet decouples. Thus the QED
coupling in (4.11) is
\begin{equation} g^2\ \sim\ -\ \frac1{\ln(m_\gamma/\Lambda_{QED})}\ .  \end{equation}
Below the photon
mass at $\mu\ll m_\gamma$ we are left with the effective string
theory (4.4)  with the coupling (4.11). We
see that $g^2_{2D}\ll1$. Thus we conclude that ANO string on the
Higgs branch does not develop large fluctuations. It represents
almost straight line connecting monopole with antimonopole. Its
string tension approximately given by its classical value $\tau$
in (1.3).

It is worth noting that the rigidity term as well as higher
corrections  cause an additional renormalization of the string
tension in (4.10) \cite{P}. Still it is known \cite{DFJ,P} that
a $\sigma$-model without $\theta$-term has no IR-fixed point
and its string tension remains finite.

This is usually considered as quite an unpleasant feature of the
theory. Suppose we try to construct reasonable fundamental
string theory starting from, say, random surfaces on a lattice
using lattice spacing as a UV-cutoff. Then in order to get the
string tension at some physical scale we need a fixed point for
the $2D$ $\sigma$-model on the worldsheet. Moreover, the string
tension should go to zero in lattice units in this fixed point
\cite{DFJ}.

In this paper we follow quite a different logic. We consider a
solitonic ANO string rather than the fundamental one. The
UV-cutoff for our string theory (4.4) is the photon mass
$m_\gamma$ ($R^{-1}_g$ more precisely), which is already a
physical parameter. The nonzero string tension given by (1.3)
ensures confinement for monopoles in the conventional setup of
this problem. Namely, for well separated $(L\gg R_g)$
heavy monopole--antimonopole pair its interaction
potential (1.4) increased with $L$.

On the other hand, if we address the problem of calculating of
the mass spectrum of hadrons on the Higgs branch then the string
approach seems to be not applicable at all.\footnote{To address
this problem with dynamical monopoles we have to go close to
Argyres--Douglas point $m=\Lambda$ or consider closed strings to
be interpreted as "glueballs".}  The reason for this is that the
distance between monopole and antimonopole in the "meson" is
$L_0\sim1/\sqrt\tau$ (at least for small spins). Thus $L_0\sim
R_g\sim1/m_\gamma$. At that short distances string is not
developed yet. We do not address problem of calculating of the
hadron mass spectrum in this paper.

Let us note also, that because $1/\sqrt\tau$ plays the role of
the correlation length for the normals on the surface, nonzero
$\tau$ means  "crumpled" string surface \cite{P,D}. This is of
course true only for open strings connecting light monopoles or
for closed strings.

We considered only boson degrees of freedom
of string in this subsection. In principle, one should add
fermions as well. However, we do not expect any supersymmetry of
our $2D$ $\sigma$-model (4.4) because our vortex is not BPS
state. Fermions produce additional renormalization of the string
tension in (4.10) which is of the same order $1/R^2_g$ as the
boson one. It is small as compared with classical $\tau$ and
does not change our main conclusions.

\subsection{Interactions of vortices}

Now we are going to study one another possibility for vortices
to develop unstability and to destabilize the Higgs branch. We
know from the condensed matter physics that vortices attract
each other and coalesce producing the unbroken vacuum in the
type one superconductor. To clarify this issue we now calculate
interactions of vortices on the Higgs branch.

To do this we use the string representation (4.4). Suppose we
have two parallel vortices of length $L$ at the separation $R$,
$R\ll L$. Then to the leading order in $g^2_{2D}$ the
contribution of these two vortices to the partition function is
\begin{eqnarray}
Z_{1,2}&=&\int Dz_1Dz_2D\rho_1D\rho_2\exp-\left\{\int d^2\sigma_1
\left[iv\rho_1+\frac{\rho^2_1}4\frac1{2\pi}\ln m_\gamma
L\right] \right.\nonumber\\
& +&  \int d^2\sigma_2\left[iv\rho_2+\frac{\rho^2_2}4\
\frac1{2\pi}\ln m_\gamma L\right]\ +\nonumber \\
&+& \left.\int d^2\sigma_1d^2\sigma_2\frac14\rho_1\rho_2
\frac1{(2\pi)^2}\frac1{[z_1(\sigma_1)-z_2(\sigma_2)]^2}
\right\},
\end{eqnarray}
where $z_{1\mu}(\sigma_1),\rho_1(\sigma_1)$ and
$z_{2\mu}(\sigma_2),\rho_2(\sigma_2)$ describes the first and
the second vortex respectively. We neglect fluctuations of
vortices in (4.13) and fix the gauge (4.5) for each vortex. Two
first terms in the exponent in (4.13) correspond to each
individual vortex while the third one describes their
interactions. Performing the integral over $(\sigma_1-\sigma_2)$
in (4.13) and integrating out $\rho_1$ and $\rho_2$ we get for
the vortex interaction potential
\begin{equation}
U_{1,2}(R)\ =\ -2\pi^2v^2TL\frac{\ln L/R}{(\ln m_\gamma L)^2}\ .
\end{equation}
The potential (4.14) shows the logarithmic attraction. It is
clear that if we consider the case $R\gg L$ then the problem
becomes three dimensional and $U_{1,2}\sim-1/R$. Note also, that
the vortex--antivortex interaction is the same as the
vortex--vortex one in the limit $m_H=0$.

Thus it is clear that vortices interact via attractive
Coulomb forces $(U\sim\ln R$ if $R\ll L$ and $U\sim-1/R$ if $R\gg
L$). However, we know that the Coulomb gas with atraction is
unstable.
 This means that if vortices condense then they attract each
other and coalesce destabilizing the Higgs branch. This is
exactly what happens in the condensed matter superconductor
when the external magnetic field exceeds its critical value.

To be more specific we can describe the interaction of the string
with the external Higgs field using eq.(4.2). It has the form
\begin{equation}
\int d^4 x i\phi(x)j(x),
\end{equation}
where $j(x)=\int d^2 \sigma \delta^{(4)}(x-z(\sigma))\rho(\sigma)$
is the "string current". The condensation of strings means
that $j$ develops nonzero VEV.  From (4.15) we see that
this would destabilize the Higgs branch.

Thus it is clear that the problem is whether vortices condense or
not.
Now we are going to argue that vortices do not condense and the
Higgs branch remains stable. First note, that open strings could
condense only if monopoles condense. However we know that
monopoles do not condense even at the Argyres--Douglas point
(where they become massless) because we have only one flavor
of monopoles and Eqs. (2.6) and (2.7) have no solutions for
monopoles. Still this argument does not rule out the possibility
of condensation of closed strings.

The more general argument is based on $N=2$ supersymmetry.
From (4.15) we expect that if
vortices condense they would generate a superpotential
\begin{equation}
\int d^2\theta\ m_0 {\wt QQ}\ ,
\end{equation}
where $m_0$ is proportional to $<j>$.
Note that the mass term is the only
 superpotential for matter fields which is allowed by
the $N=2$ supersymmetry  ( if gauge fields
are absent, see, for example, review
\cite{S}). However, (4.16) does not destabilize the
Higgs branch. It just produce a shift in the bare mass
parameter $m$.

This argument has even more general nature. If vortices
destabilize the Higgs branch we expect them to generate some
superpotential on it. As soon as no non-trivial superpotential is
allowed without breaking $N=2$  supersymmetry we conclude that the
Higgs branch is stable. This means that vortices do survive in the
quantum theory giving rise to the confining potential (1.4)
between heavy well separated monopoles on the Higgs branch.

\section{Conclusions}

In this paper we studied vortices on the Higgs branch of the
Seiberg--Witten model. We showed that vortices of finite length
exist classically and produce the confining potential (1.4) for
heavy monopoles. This confining potential is not linear in $L$
and we can use Wilson loop as an order parameter to distinguish
this new confining phase specific for Higgs branches in
supersymmetric theories.

Then we studied the stability of vortices in the quantum theory.
First, we showed that vortices do not grow in their transverse
size $R_g$. Second, we studied vortices in the string
representation and showed that ANO strings do not develop large
fluctuations and their string tension remains nonzero in the
quantum theory.

Next we ruled out the possibility for vortices to condense and
form the Coulomb gas with attraction which would destabilize the
Higgs branch. Our conclusion is that vortices survive in the
quantum theory producing a new type of confining phase for
monopoles with the potential (1.4).

We have not addressed the problem of calculating of the hadron
mass spectrum on the Higgs branch. As we have shown the string
approach is not acceptable for this purpose. Still it is quite
plausible to suggest on the general grounds that the lightest
monopole--antimonopole "meson" becomes massless at the
Argyres--Douglas point $m=\Lambda$. To see the reason for this
let us recall that the root of a Higgs branch at $m=\Lambda$
corresponds to conformal field theory with non-trivial anomalous
dimensions \cite{APSW}. If all "hadrons" built of monopoles are
massive at this point we could integrate them out together with
photon multiplet at scales $\mu^2\ll\langle\bar qq\rangle$. Then
we would get a $\sigma$-model for charges $q_\alpha$ with the
flat metric (the Higgs branch admits only unique hyper--Kahler
metric). Then nothing prevents us from taking the limit
$\langle\bar qq\rangle\to0$, in which we would get trivial fixed
point. This is correct for all values of mass parameter except
Argyres--Douglas points $m=\pm\Lambda$, where we should recover
non-trivial conformal field theory.

The possible resolution of this puzzle is that there is a
massless monopole (dyon) "meson" at $m=\Lambda(-\Lambda)$ (see
also \cite{Y}).  Still the calculation of the "hadron" mass
spectrum on the Higgs branch remains to be an open problem.

\subsection*{Acknowledgments}

The author is grateful to A.~Gorsky, A.~Marshakov, V.~Rubakov,
and A.~Schwimmer for useful discussions. In particular, the
author would like to thank A.~Vainshtein for numerous and very
illuminating discussions. Also the author is grateful to the
Theoretical Physics group at the University of
Minnesota where a part of this work has been done for
hospitality. This work is supported by Russian Foundation for
Basic Research under grand No.~99-02-16576.

\end{document}